\documentclass[fleqn,12pt,twoside]{article}
\usepackage[headings]{espcrc1}

\readRCS
$Id: espcrc1.tex,v 1.2 2004/02/24 11:22:11 spepping Exp $
\usepackage{graphicx}
\usepackage[figuresright]{rotating}

\newcommand{\AmS}{{\protect\the\textfont2
  A\kern-.1667em\lower.5ex\hbox{M}\kern-.125emS}}

\def\empile#1\over#2{\mathrel{\mathop{\kern 0pt#1}\limits_{#2}}}
\def\bs{\boldsymbol}

\def\q{{\boldsymbol q}}

\def\k{{\boldsymbol k}}

\def\x{{\boldsymbol x}}
\def\y{{\boldsymbol y}}

\def\r{{\boldsymbol r}}
\def\z{{\boldsymbol z}}

\def\b{{\boldsymbol b}}
\def\u{{\boldsymbol u}}

\usepackage{amssymb}
\usepackage{amsmath}
\usepackage{graphicx}
\newcommand{\beq}{\begin{eqnarray}}
\newcommand{\eeq}{\end{eqnarray}}

\long\def\comment#1{ }    
\newcommand{\be}{\begin{equation}}
\newcommand{\ee}{\end{equation}}

\title{A unified description of initial and final state interactions in heavy ion collisions}

\author{Y. Mehtar-Tani\address[MCSD]{ Laboratoire de Physique Th\'eorique\\
  Universit\'e Paris Sud, B\^at. 210\\
  91405, Orsay cedex}%
,}

\begin{document}

\maketitle

\begin{abstract}
We have investigated the gauge field of a fast moving projectile passing through a dense medium. We provide a simple and unified derivation, in light cone gauge, of the gluon production cross-section in proton-nucleus (initial state interactions), and the radiative gluon spectrum of a hard parton produced in a nucleus-nucleus collision (final state interactions). Finally, we discuss the validity of the eikonal approximation in proton-nucleus collisions at RHIC energies.  
\end{abstract}

\section{Introduction}
Heavy ion collisions at high energy have stimulated a lot of theoretical progresses in the understanding of QCD at extreme densities and energies. Two important orientations have been taken especially after RHIC experiments have started. The first one, consists on searching for the Quark-Gluon-Plasma (QGP) in nucleus-nucleus collisions. The QGP is a thermalized hot and dense medium made of deconfined quark and gluon matter. This medium would be formed soon after a high energy nucleus-nucleus collisions. RHIC experiments has proved the formation of a such dense medium, the question of whether it is thermalized is still open. Indeed, a strong suppression of produced jets has been observed as compared to the proton-proton collisions (see for instance \cite{star}). This suppression is explained by energy loss of hard partons produced initially in the collision . It is believed that most of the energy is lost by radiating gluons off the hard parton when passing through the dense medium. This picture is known as the BDMPS-Z-W description for radiative energy loss \cite{BDMPS1,BDMPS3,Zak1,Wied1,Wied2}. Let us give in a few words the main ingredients. \\
The hadron production cross section is modified by parton interactions in the medium as follows 

\be
\frac{d\sigma^{med}(p_\perp)}{dyd^2{\boldsymbol p}_\perp}=\int d\varepsilon P(\varepsilon) 
\frac{d\sigma^{vac}(p_\perp+\varepsilon)}{dyd^2{\boldsymbol p}_\perp},
\ee
where $P(\varepsilon)$ is the density probability for medium induced parton energy loss $\varepsilon$. $\sigma^{vac}$ is the hadron production cross section in vacuum i.e. in the absence of the medium. It is well described by perturbative QCD at high $p_t$.  \\
In the approximation of independent gluon emissions, the probability density can be taken to be a Poissonian distribution \cite{BDMPS3}
\be
P(\varepsilon)=\sum_{n=0}^\infty \frac{1}{n!}\left[\prod_{i=1}^n \int d\omega_i\frac{dI (\omega_i)}{d\omega}\right]\delta(\varepsilon-\sum_{i=1}^n \omega_i)e^{-\int d\omega
 \frac{dI}{d\omega}}, 
\ee
where the radiative gluon spectrum $I$, is related to the average number of radiated gluons off the hard parton: 
\be
\omega\frac{ dI}{d\omega}=\int d^2\q_\perp\omega\frac{ d\langle N\rangle}{d\omega d^2\q_\perp}.
\ee
In the BDMPS-Z formalism, one assumes that the hard parton and the radiated gluon scatter coherently in the medium, i.e. they suffer multiple scatterings with gluons and quarks present in the medium. \\
The second orientation taken in the study of heavy ion collisions at high energy, is the description of initial state interactions. Namely, the description of the nuclear wave function at small $x$. Proton-nucleus collisions allows one to probe directly the nuclear wave function since in that case final state interactions are absent (see for instance \cite{BRAHMS1}). The theory of the Color Glass Condensate (CGC) \cite{McLerV1,IancuLM1,FerreILM1,JalilKLW1,JalilKMW1,Balit1,Kovch3,KovneM1,KovneMW3} has been developed to describe the nuclear wave function in a regime where the nucleus contains a large number of overlapping gluons. In this saturation regime \cite{GriboLR1,MuellQ1}, gluons dominate the dynamics. Thus, gluon production in proton-nucleus collisions is a convenient observable to reach this regime delimited by the saturation scale $q_\perp\lesssim Q_s(x)$ where  $q_\perp$ is the transverse momentum of the produced gluon. \\The proton can be seen, in a first approximation, as three valence quarks which scatter in the dense nucleus and radiate a gluon; the quark-gluon system suffers multiple rescatterings inside the nucleus. The gluon gets, by this mean, a large transverse momentum transfer of order of the saturation scale. The gluon production cross section in proton-nucleus collisions is related to the average number of produced gluons of transverse momentum $\q_\perp$ and rapidity $y$ ($dy\equiv d\omega/\omega$) :
\be
\frac{d\sigma^{pA}}{dyd^2\q_\perp d^2\b_\perp } \equiv\omega \frac{d\langle N\rangle}{d\omega d^2\q_\perp},\nonumber
\ee
$\b_\perp$ is the impact parameter of the collision. \\
We will show that both the radiative gluon spectrum in nucleus-nucleus collisions and the gluon production cross section in proton-nucleus collisions can be described within the same formalism. The basic idea is to calculate the classical gluon field of a fast moving projectile passing through a dense medium described by static classical color sources $\rho_0(x^+,\x_\perp)$ in the light cone gauge of the projectile $A^+=0$:\\
i) In nucleus-nucleus collisions the medium is the dense medium produced soon after the collision and the fast projectile is the hard parton.\\
ii) In proton-nucleus collisions the medium is the nucleus and the fast projectile is the proton (assumed to be made of three valence quarks).\\
The differential average number of produced gluons is given by the formula:
       
       \be
       \omega\frac{d\langle N \rangle}{d\omega d^2\q_\perp}=\frac{1}{16\pi^3} \langle \sum_\lambda|\int d^4x \square_x \delta A_\mu(x) \epsilon^\mu_{(\lambda)}(\q) e^{ix.q}|^2   \rangle   \label{eq:dN}
       \ee
       where $\epsilon^\mu_{(\lambda)}$ is the polarization vector of the radiated gluon, and $q\equiv(q^+=\omega,q^-=\q_\perp^2/2\omega,\q_\perp)$ its momentum. The average $\langle N\rangle$ means that we average over the medium sources $\rho_0$ by defining a statistical weight ${\cal W}$ :
\begin{equation}
    \langle N\rangle\equiv\int {\cal D}[\rho_0] {\cal W}[\rho_0]N[\rho_0].
    \end{equation}
 
\section{The gauge field of a fast moving parton (or hadron) passing through a dense medium}
\begin{figure}[ht]
\begin{center}
\resizebox*{!}{4cm}{\includegraphics{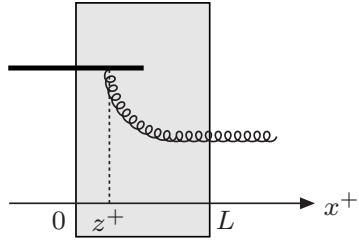}}
\caption{A schematic representation of eq. (\ref{eq:field}). The fast projectile (thick line) passes through the medium of thickness $L$ and emits a gluon at the time $z^+$. The gluon emission could also occur outside the medium, at $z^+<0$ or $z^+>L$.  }\label{fig1}
\end{center} 
\end{figure}

We consider a massless parton (or a hadron described as a collection of partons) moving with the velocity of light, in the $x^+$ direction. At some time it passes through a dense static medium of size $L$. The medium is described by the following gauge field, in the medium rest frame, 
\begin{equation}
{\tilde A}_0^\mu=\delta^{0\mu}\frac{1}{\square}{\tilde \rho}_0(x^3,\x_\perp) \delta^{\mu 0}.
\end{equation}
Knowing the simplicity of solving the Yang-Mills equations in the light-cone gauge of the parton $A^+=0$ \cite{GM}, we perform a boost of velocity $\beta\sim-1$ (the cross-section is Lorentz invariant) which affects only the medium, namely the parton remains in the $x^+$ direction and the medium is pushed very close to the light-cone in the $x^-$ direction. \\    
In the boosted frame the medium field can be checked to be \footnote{For any gauge choice for the medium field in its rest frame the $A^+$ component is suppressed by the boost.} \cite{BGV1}:
  \begin{equation}
A^\mu_0=-\delta^{-\mu}
\frac{1}{{\bs\partial}_\perp^2}\rho_0(x^+,\x_\perp)\;\label{eq:A-}.
\end{equation} 
We assume that when the parton traverses the medium, it induces a perturbation $\delta A^\mu$ of the strong medium field $A_0$ (linear response):
\begin{equation}
A^\mu=A^\mu_0+\delta A^\mu\;.
\end{equation}
Similarly for the conserved current
 \begin{equation}
J^\mu=J^\mu_0+\delta J^\mu\;.
\end{equation}
Keeping only terms in the Yang-Mills equations which are linear in the fluctuation  $\delta A^\mu$:
\begin{eqnarray}
&&
-\partial^+(\partial_\mu \delta A^\mu)=\delta J^+,
\nonumber\\
&&
\square \delta A^i-2ig[A_0^-,\partial^+\delta A^i]=\partial^i(\partial_\mu \delta A^\mu)
\; .\label{eq:YM}
\end{eqnarray}
The parton current obeys the conservation relation $[D_\mu,J^\mu]=0$. With the initial condition $\delta J^+(x^+=t_0)=\delta(x^-)\rho(\x_\perp)$, where $t_0$ is the source production time, namely the production time of the hard parton in nucleus-nucleus collisions. In writing this current, we assume that the hard projectile propagates in the $x^+$ direction of the light cone; therefore, its interaction with the medium is eikonal: it only gets a color precession when passing through the medium. The solution for the current reads
\begin{equation}
\delta J^\mu=\delta^{\mu+}U(x^+,t_0,\x_\perp) \delta(x^-)\rho(\x_\perp)\theta(x^+-t_0)-\delta^{\mu-}\theta(x^-)\rho(\x_\perp)\delta(x^+-t_0)\; \label{current}.
\end{equation} 
where $U$ is a Wilson line in the adjoint representation of the
gauge group:
\begin{equation}
U(x^+,t_0,\x_\perp)\equiv {\cal T}_+\, \exp \left[ig\int_{t_0}^{x^+}
dz^+\; A_0^{-a}(z^+,\x_\perp)T_a\right]\;.
\end{equation}
The second term in (\ref{current}) corresponds to an antiparticle moving in the direction $x^-$, it plays no role but it is necessary to get a conserved current. The solution of the transverse field in (\ref{eq:YM}) is easily obtained (The $-$ component is not relevant for the spectrum) and reads 
\begin{equation}
\delta A^i(x)=-\int d^4z\theta(z^-)G(x,z)\partial^i(U(z^+,t_0,\z_\perp)\rho(\z_\perp))\theta(z^+-t_0)\;.
\label{eq:field}
\end{equation}
The gluon propagator $G$ is the retarded Green's function obeying the equation of motion:
\begin{equation}
(\square_x -2ig(A_0^-.T)\partial_x^+)G(x,y)=\delta(x-y)\;.\label{eq:Green}
\end{equation}
 The propagation of the emitted gluon, contrary to the hard projectile, is not necessarily eikonal.
Eq. (\ref{eq:field}) has a simple diagrammatic representation, shown in fig. (\ref{fig1}). The color precession of the source before the gluon emission is accounted for by the $U$ that multiplies $\rho$; while the rescatterings of the gluon after it is emitted are hidden in the Green's function $G$.

\section{Induced radiative spectrum in nucleus-nucleus collisions} 

We assume that the fast projectile is a hard parton produced in a nucleus-nucleus collision at $t_0$, and take the origin of times when the parton enters the medium. In coordinate space, (for $x^+>L$) the gluon field (\ref{eq:field}), amputated of its final free propagator, reads    
\begin{eqnarray}
 \square_x \delta A^i(x)&=&-\theta(x^-)\partial^i_x(U(x^+,t_0;\x_\perp)\rho(\x_\perp))-\int d^4z\theta(z^-)\theta(L-z^+)\theta(z^+-t_0)\delta(x^+-L)\nonumber\\
&&\times2\partial^+_xG(x,z)\partial^i_z(U(z^+,t_0;\z_\perp)\rho(\z_\perp)).\;\label{eq:ampfield}
\end{eqnarray}
The invariance by translation, with respect to the $-$ variables, of this Green's function is due to the fact that the medium field $A^-_0$ is independent of $x^-$, this can be seen in eq. (\ref{eq:Green}). This invariance implies that $G$ depends on the coordinates only via $x^--y^-$.\\
 It is useful to introduce a new Green's function, defined as 
\begin{equation}
{\cal G}_\omega(x^+,\x_\perp;y^+,\y_\perp)=2\int dl^- \partial_x^+G(x^+,\x_\perp;y^+,\y_\perp;l^-=x^--y^-)e^{il^-\omega}. 
\end{equation}
For a single parton localized at $\x_\perp={\bf 0}_\perp$ we have 
\begin{equation}
\rho(\x_\perp)=\sqrt{\frac{g^2C_R}{N_c^2-1}}\delta(\x_\perp),
  \end{equation}
where $R=A$ for a gluon and $R=F$ for a quark.\\
To get the average number of the produced gluons, we square the
amplitude, sum over the polarization vectors. From (\ref{eq:dN}), the gluon spectrum reads  
\begin{eqnarray}
&&\omega\frac{d\langle N \rangle}{d\omega d^2\q_\perp}= \frac{g^2C_R}{(2\pi)^3(N_c^2-1)\omega}\Re e\int d^2\x_\perp\int d^2\y_\perp  e^{-i(\x_\perp-\y_\perp).\q_\perp}\int_{t_0}^L dz^+
\Big[\frac{1}{\omega}\int_{t_0}^{z^+}dz'^+\nonumber\\
&&\times
\left<{\bf tr}U(z^+,z'^+,\z_\perp)\partial^i_{z'}{\cal
G}^{\dag}_\omega(L,\y_\perp;z'^+,\z'_\perp)\partial^i_z{\cal
G}_\omega(L,\x_\perp;z^+,\z_\perp=0)\right>\vert_{\z_\perp=\z'_\perp}\nonumber\\ &&-2\delta(\y_\perp)\frac{q^i}{\q_\perp^2}\left<{\bf tr }U^\dag(L,z^+,\y_\perp)\partial^i_y{\cal
G}_\omega(L,\x_\perp;z^+,\y_\perp)\right> \Big]+\frac{2g^2C_R}{(2\pi)^3\q_\perp^2}.
\label{eq:dI}
\end{eqnarray}
The first term in (\ref{eq:dI}) corresponds to the probability of producing the gluon inside the medium, whereas the second term corresponds to the interference between the amplitudes for producing the gluon outside and inside the medium. The last term is the probability of radiating a gluon in the vacuum, it has to be removed to get the medium induced gluon spectrum. This formula is quite general, and we will show that it leads to the well known BDMPS-Z-W spectrum, in the case of uncorrelated sources. This approximation assumes that the scattering centers at different times in the medium are independent\footnote{This approximation holds when the range of one scattering center is much smaller than the mean free path of the radiated gluon and of the hard parton inside the medium.}, therefore the medium sources can be treated as Gaussian, namely, that amounts to choose the statistical weight ${\cal W}$ as follows
\begin{equation}
{\cal W}[\rho_0]=\int {\cal D}[\rho_0]\exp\left\lbrace -\frac{1}{2}\int dx^+d^2\x_\perp\frac{\rho^a_0(x^+,\x_\perp)\rho^a_0(x^+,\x_\perp)}{n(x^+)}\right\rbrace ,
\end{equation}
where $n(x^+)$ is the medium scattering center density at the time $x^+$. In this approximation, the two-point function reads
\begin {equation}
\left<{\bf tr } U(z^+,z'^+,\x_\perp)U^\dag(z^+,z'^+,\y_\perp)\right>=\exp\left[-\frac{1}{2}\int^{z^+}_{z'^+}d\xi n(\xi) \sigma(\x_\perp-\y_\perp)\right]\label{eq:UU}
\end {equation}
Introducing the quantity 
\be
{\cal K}_\omega(z'^+,\y_\perp-\z_\perp;z^+,\x_\perp-\z_\perp)=\int{\cal
D}\r_\perp(\xi)\exp\left[\int^{z^+}_{z'^+}d\xi (\frac{i\omega}{2}
{\dot\r}_\perp^2(\xi)-\frac{1}{2}n(\xi) \sigma(\r_\perp))\right],
\ee
where $\r_\perp(z'^+)=\y_\perp-\z_\perp$ and $\r_\perp(z^+)=\x_\perp-\z_\perp$.\\
We recover the well-known induced radiative gluon spectrum \cite{BDMPS3,Wied2}.
\begin{eqnarray}
&&\omega\frac{ d\langle N\rangle}{d\omega d^2 \q_\perp}=\frac{\alpha_sC_R}{(2\pi)^2\omega}2\Re e\int_{t_0}^L dz^+\int d^2 \u_\perp e^{-i\q_\perp.\u_\perp}\nonumber\\
&&\times\Big[ \frac{1}{\omega}\int_{t_0}^{z+} dz'^+e^{-\frac{1}{2}\int_{z'^+}^L  d\xi n(\xi)\sigma(\u_\perp)} {\bs\partial}_{\perp y}. {\bs\partial}_{\perp u}{\cal K}_\omega(z^+,\u_\perp;z'^+,\y_\perp=0) \nonumber\\
&&-2\frac{\q_\perp}{\q_\perp^2}. {\bs\partial}_{\perp y}{\cal K}_\omega(L,\u_\perp;z^+,\y_\perp=0)\Big].
\end{eqnarray}
 
\section{Gluon production in proton-nucleus collisions}
\begin{figure}[hbtp]
\begin{tabular}{c  c }

\resizebox*{!}{4cm}{\includegraphics{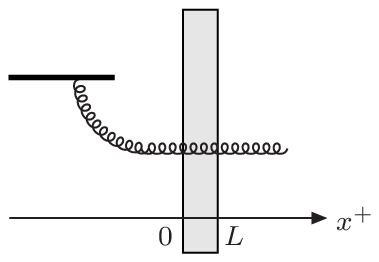}} &\resizebox*{!}{4cm}{\includegraphics{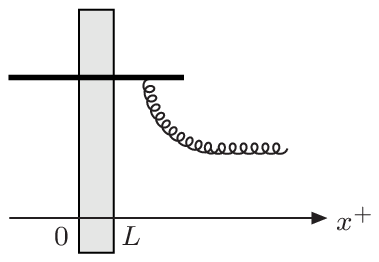}} \\
      \qquad (a)& \qquad(b)
\end{tabular} \caption{The two diagrams contributing to gluon production in proton-nucleus collisions in the limit $L\rightarrow 0$. In this limit the gluon is eikonal when passing through the nucleus, and the gluon emission inside the nucleus is neglected. 
} \label{fig2}
\end{figure}

At high energy the nucleus is Lorentz contracted, so that we take the limit $L\rightarrow 0$, and put $t_0=-\infty$. As a consequence the produced gluon is eikonal during the interaction time $L$, thus, the gluon field (\ref{eq:field}), when amputated of its final free propagator, reduces to \cite{GM}
\begin{eqnarray}
\square \delta A^i(x)&=&
2\delta(x^+)\delta(x^-)(U-1)\frac{\partial^i}{{\bs\partial}^2_\perp}
\rho(\x_\perp)
\nonumber\\
&-&
\theta(x^-)\theta(-x^+) \partial^i \rho(\x_\perp)
-\theta(x^-)\theta(x^+) \partial^i(U\rho(\x_\perp))\; ,\label{eq:fieldpA}
\end{eqnarray}
where $U\equiv U(+\infty,-\infty;\x_\perp)$. Then the Fourier transform gives
\begin{eqnarray}
-q^2 \delta A^i(q)&=&
-q^2 A_{\rm proton}^i(q)+
2i\int \frac{d^2\k_{1\perp}}{(2\pi)^2}
\left[\frac{q^i}{2(q^++i\varepsilon)(q^-+i\varepsilon)}
-\frac{k_1^i}{k_{1\perp}^2}
\right]\nonumber\\
&&\times\rho(\k_{1\perp})
\left[U(\k_{2\perp})-(2\pi)^2\delta(\k_{2\perp})\right]\;,\label{eq:fieldpA2}
\end{eqnarray}
where $\k_{2\perp}\equiv \q_\perp-\k_{1\perp}$ and
$A_{\rm proton}^i(q)$ is the Fourier transform of the gauge field
of a proton alone, i.e. the Fourier transform of
eq.~(\ref{eq:fieldpA}) taking $U=1$. The two terms in (\ref{eq:fieldpA2}) are illustrated in fig. (\ref{fig2}). This expression
leads to the standard result for gluon production in
proton-nucleus collisions \cite{GM,BGV1,KovnW,KovchM3,DumitM1,KovchT1}.
\section{Validity of the eikonal approximation in proton-nucleus collisions at RHIC}
The center of mass energy per nucleon at RHIC is  200 GeV. It corresponds to a Lorentz contraction factor of about $\gamma=\sqrt{s}/2m_p\simeq 100$. For a nuclear radius $R_A\simeq 6.5$ fm, we end up with the estimate $L\simeq 0.5$ GeV$^{-1}$.
For the eikonal approximation to be valid, the gluon production time has to be much bigger than the maximum interaction time:
\begin{equation}
t_{prod}\sim\frac{\omega}{\q_\perp^2}\gg L . 
\end{equation}
At mid-rapidity, at RHIC, $ \omega\sim q_\perp\sim Q_s\sim$ 1GeV. We get $t_{prod}$ of order of $L$.  Therefore, one probably has  to go beyond the eikonal approximation for the gluon propagator to describe mi-rapidity data at RHIC. \\
The corrections to the eikonal approximation for the gluon propagation inside the medium are taken into account in formula (\ref{eq:dI}). Even if (\ref{eq:dI}) has been derived in the framework of final state interactions in nucleus-nucleus, it is applicable for gluon production in proton-nucleus collisions (assuming the medium to be a nucleus and the projectile to be a proton, and taking the projectile production time $t_0=-\infty$). This is a possible phenomenological application for mid-rapidity hadron production at RHIC. \\
At forward rapidity, $\omega\sim q_\perp e^\eta$, therefore, $\omega$ is enhanced by a large factor (exponential of the rapidity) leading to a much larger gluon production time compatible with the eikonal approximation.

\end{document}